\documentstyle[aps]{revtex}
\input psfig.sty

\begin{document}

\title{Dissipation of angular momentum in light heavy ion collision}

\author{C. Bhattacharya, S. Bhattacharya, T. Bhattacharjee, A. Dey, 
S. Kundu, S. R. Banerjee,  P. Das, S. K. Basu, K. Krishan}

\address{ Variable Energy Cyclotron Centre, 1/AF Bidhan Nagar,
Kolkata - 700 064, India}


\pagenumbering{arabic}
\maketitle

\begin{abstract}

The  inclusive  energy  distributions  of  fragments  (  4$\leq$Z$\leq$7)
emitted in the reactions $^{16}$O (116 MeV) + $^{27}$Al, $^{28}$Si, 
$^{20}$Ne (145 MeV) + $^{27}$Al, $^{59}$Co have
been  measured  in  the  angular  range  $\theta_{lab}  $=  10$^\circ$  -
65$^\circ$. The respective fusion-fission and  deep  inelastic 
contributions  have  been decomposed from   the experimental 
fragment energy spectra. The angular mometum dissipations
in fully damped deep inelastic collisions have been estimated assming 
exit channel configuration similar to those for fusion-fission process. 
It has been found that, the angular momentum dissipations are more than
those predicted by the empirical sticking limit in all cases. The deviation
is found to increase with increasing charge transfer (lighter fragments). 
Qualitatively, this may be due to stronger friction in the exit channel. 
Moreover, for the heavier system $^{20}$Ne +  $^{59}$Co, the 
overall magnitude of deviation is less as compared to those for the 
lighter systems, {\it i.e.},  $^{16}$O + $^{27}$Al, $^{28}$Si, $^{20}$Ne + 
$^{27}$Al. This may be due to lesser overlap in time scales of fusion
and deep inelastic time scales for heavier systems.

\end{abstract}


\section{INTRODUCTION}
\label{sec:int}
Several experimental studies have been made in the recent years to
understand the reaction mechanism of fragment emission in light heavy
ion collisions \cite{bh,su,sa99,ber,be1,be2,be3,be4,ba,se,sh1,sh2,na,eg,co}  
at low bombarding energies ($\lesssim$ 10 MeV/nucleon). The fragments 
(mostly binary in nature at these energies) are emitted  with different
degree of dissipation of the entrance channel kinetic energy between the
two colliding ions - ranging from quasi elastic to deep inelastic (DI) to the fully
relaxed fusion-fission (FF) processes. Thus the fragments carry the signatures 
of nuclear dissipation, which, if deciphered, may bring out valuable 
informations on the nature of nuclear dissipation.

In addition to kinetic energy dissipation, dissipative heavy ion collision 
processes also result in significant dissipation of relative angular momentum
in the entrance channel. Phenomenologically, the kinetic energy dissipation 
originates from friction (radial and tangential) between the surfaces of the rotating 
dinuclear system; on the other hand, angular momentum 
dissipation is decided solely by the tangential component of the friction, and
the magnitude of dissipation is expected  to lie between two limits (rolling 
and sticking). However, very large dissipation of relative angular momentum 
in excess of the sticking limit predictions has also been reported in the
literature \cite{sh2}. This anomaly, as pointed out  by several authors 
\cite{se,eg,co2,bet,bra}, is due to the ambiguity in the determination of 
the magnitude of angular 
momentum dissipation (and vis-a-vis the rotational contribution to the fragment
kinetic energy). Estimation of the angular momentum in the exit channel is strongly 
dependent  on  another poorly known factor, i.e., the scission
configuration of the rotating dinuclear system.
This apparently hinted at the incompleteness of
our understanding of the dynamics of nuclear dissipation process, which  
prompted  us to make a systematic study  of angular momentum dissipation
in light nuclear systems where Coulomb and rotational contributions to the 
fragment kinetic energies are comparable.  

It is clear from the above that an independent estimation of the scission 
configuration is necessary to make a proper estimate of the angular momentum
transfer. Generally, it is estimated from the total kinetic energy of the 
rotating dinuclear system, $E_k$, which is given by,

\begin{equation}
E_k = V_N(d) + f^2 {\hbar^2 l_i (l_i+1) \over 2 \mu d^2} ,
\label{eq:ef}
\end{equation}

where $V_N(d)$ is the contribution from Coulomb and nuclear forces at dinuclear
separation distance $d$, $\mu$ is the reduced mass of the dinuclear configuration,
$l_i$ is the relative angular momentum in the entrance channel and $f$ is the numerical 
factor denoting the fraction of the angular mometum transferred depending on the type 
of frictional force. Since it is not possible to determine both $f$ and $d$ by solving
Eq.~\ref{eq:ef}, in the earlier works, one of them was estimated phenomenologically 
(usually for  $f$,  its value corresponding to sticking limit was taken) and the other was estimated
from the experimental fragment kinetic energy data using Eq.~\ref{eq:ef}. It is however well
known (see, for example, Ref.~\cite{bh} and references therein) that, apart from dissipative
collision process, fusion-fission process also contributes significantly in the 
fragment emission scenario. Thus, it is required to estimate and separate out the FF part of the
fragment energy spectra in order to extract the kinetic energy distribution of the DI part of the 
rotating dinuclear system. In the present work, we have studied fragment emission
from $^{16}$O (116 MeV) + $^{27}$Al, $^{28}$Si, $^{20}$Ne (145 MeV) + $^{27}$Al, $^{59}$Co
and report on how angular momentum dissipation can be estimated from the FF and DI components 
extracted by  nonlinear optimisation
procedure using multiple Gaussians \cite{bh}. Some parts of the 
$^{16}$O (116 MeV) + $^{27}$Al data have already been published in Ref.~\cite{bh}, where it
has been shown that the FF component is quite competitive, in agreement with 
the previous work \cite{sz97}.

The paper has been organised as follows. Experimental details and results have been 
described in Sec.~\ref{sec:exp}. Discusssions of the results have been given in 
Sec.~\ref{sec:dis}. Finally, the summary and concluding remarks have been given in
Sec.~\ref{sec:sum}. 

\section{EXPERIMENTS AND  RESULTS}
\label{sec:exp}
The  experiment  was  performed  using $116$ MeV $^{16}$O$^{5+}$ and 
$145$ MeV $^{20}$Ne$^{6+}$ ion beams
from the  Variable  Energy  Cyclotron  at  Kolkata. 
Self-supporting targets of 420  $\mu$g/cm$^2$  $^{27}$Al, 
$\sim$1 mg/cm$^2$ $^{28}$Si and $\sim$2 mg/cm$^2$ $^{59}$Co 
were used in the experiment.   The
fragments  were  detected  using  three  solid state (Si(SB)) telescopes (
$\sim$ 12$\mu m$ $\Delta$E, 300$\mu m$ E) mounted on one arm of the  91.5
cm scattering chamber. Typical solid angle subtended by each detector was
$\sim$0.3  msr.  A monitor detector ($\sim$300$\mu m$ Si(SB)) was placed on
the other arm of the scattering chamber for  normalisation  purpose.  The
telescopes  were calibrated using elastically scattered $^{16}$O and $^{20}$Ne 
ions from Au target and $\alpha$-particles from ($^{229}$Th-$\alpha$) source. 
Energy losses of the incoming beam as well as the outgoing fragments in
the target have been properly taken care of.

\begin{figure}


\centering
{\psfig{figure=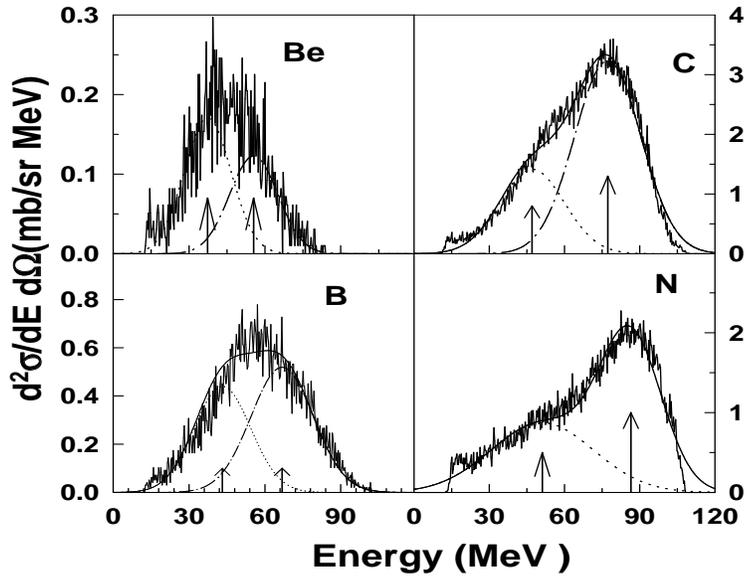,width=10.5cm,height=12.0cm}}


\caption{  Typical energy  spectra  of different fragments obtained at 15$^\circ$
for the $^{16}$O+$^{27}$Al reaction.  Dotted, dash-dot and solid
curves  represent contributions of FF, DI and  their sum (FF+DI), respectively. 
Left and  right  arrows  correspond  to   the   centroids   of   FF   and   DI
energy distributions, respectively.} 
\label{fig1} 
\end{figure}

Inclusive  energy  distributions  for various fragments (4$\leq Z \leq$7)
were measured in the angular  range  10$^\circ$-65$^\circ$.  Typical  energy
spectra  of  the  fragments (4$\leq Z \leq$7)  emitted in the reaction
$^{16}$O (116 MeV) + $^{27}$Al have been shown in
Fig.~\ref{fig1} for $\theta_{lab} = 15 ^{\circ}$. The  systematic  errors  in  the
data,  arising from the uncertainties in the measurements of solid angle,
target thickness and the  calibration  of  current  digitizer  have  been
estimated to be $\approx$ 10\%. 

\begin{figure}


\centering
{\psfig{figure=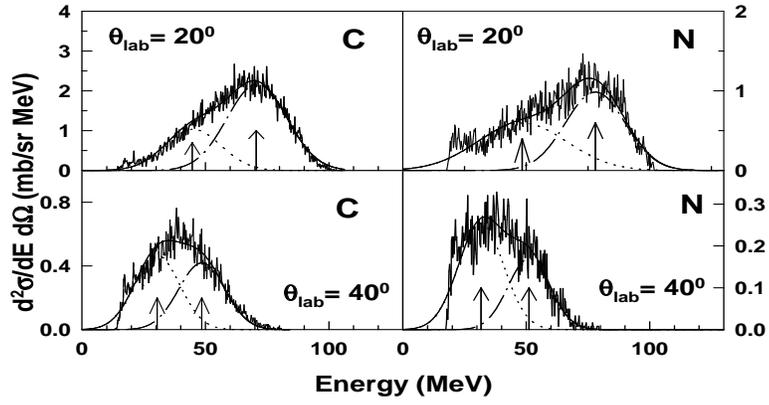,width=10.5cm,height=12.0cm}}


\caption{  Energy  spectra  of Carbon and Nitrogen fragments  at 20$^\circ$
and 40$^\circ$ for the $^{16}$O+$^{27}$Al reaction.  Dotted, dash-dot and solid
curves  represent contributions of FF, DI and  their sum (FF+DI), respectively. 
Left and  right  arrows  correspond  to   the   centroids   of   FF   and   DI
energy distributions, respectively.} 
\label{fig2a} 
\end{figure}

\begin{figure}


\centering
{\psfig{figure=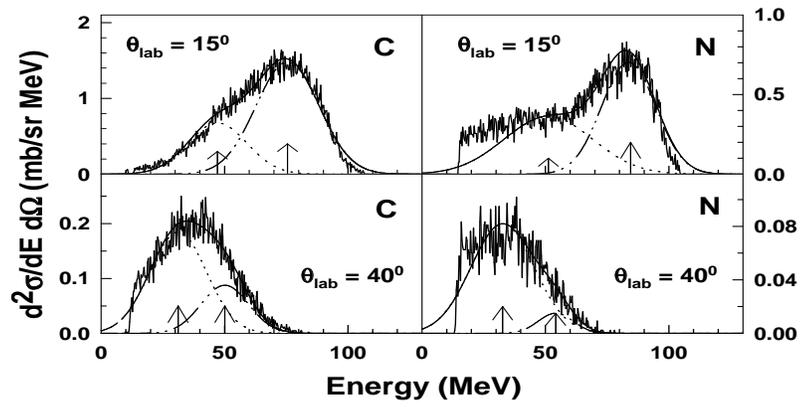,width=10.5cm,height=12.0cm}}


\caption{  Same as Fig.~\ref{fig2a} at 15$^\circ$
and 40$^\circ$ for the $^{16}$O+$^{28}$Si reaction.  } 
\label{fig2b} 
\end{figure}

\begin{figure}


\centering
{\psfig{figure=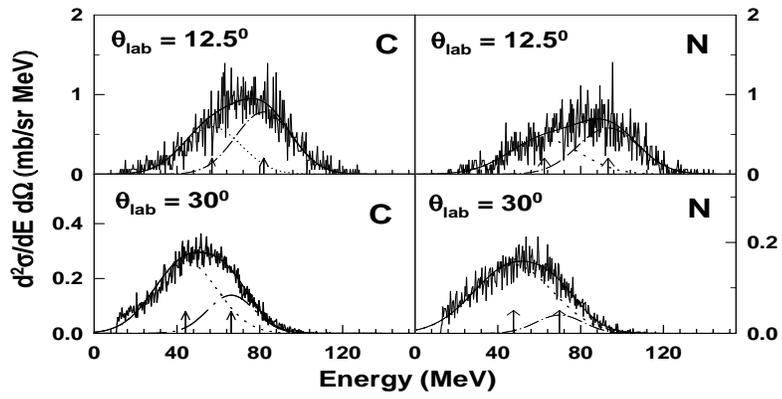,width=10.5cm,height=12.0cm}}


\caption{  Same as Fig.~\ref{fig2a} at 12.5$^\circ$
and 30$^\circ$ for the $^{20}$Ne+$^{27}$Al reaction.  } 
\label{fig2c} 
\end{figure}

\begin{figure}


\centering
{\psfig{figure=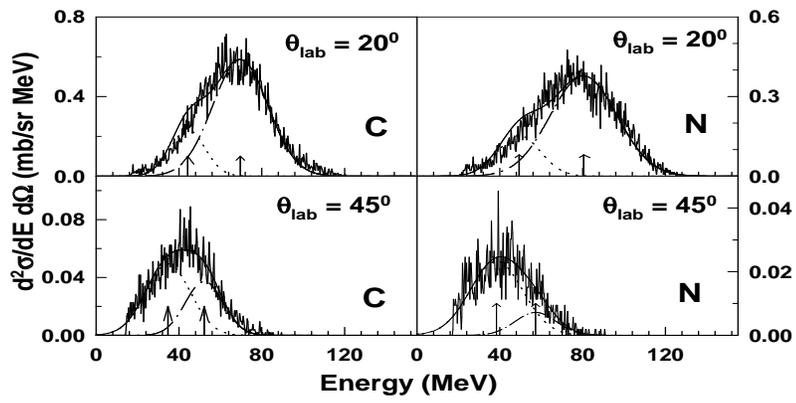,width=10.5cm,height=12.0cm}}


\caption{  Same as Fig.~\ref{fig2a} at 20$^\circ$
and 45$^\circ$ for the $^{20}$Ne+$^{59}$Co reaction.  } 
\label{fig2d} 
\end{figure}

\paragraph{Decomposition of FF and DI components : }
The contributions of fusion-fission and deep inelastic (DI) components are
estimated by fitting the measured spectra with  Gaussian functions as per the 
procedure laid down in Ref.~\cite{bh}. 
The energy spectra of different fragments at each
angle  have been fitted with two Gaussian functions in two steps.
In the first step, the FF contributions have been obtained by fitting the
energy distributions with a Gaussian  having  centroid  at  the  energies
obtained  from  Viola  systematics  \cite{vi}, adapted for light nuclear systems
\cite{be96},  of total kinetic
energies(TKE) of mass-symmetric  fission  fragments  duly  corrected  for
asymmetric  factor \cite{be1}. The width of the Gaussian was obtained
by fitting the lower energy tail  of  the  spectra.  The  FF  component of the energy
spectrum thus obtained was then substracted from the full energy spectrum.
In the next step, the DI component was obtained by fitting the substracted
energy spectra with a second Gaussian.
This is illustrated for $^{16}$O (116 MeV) + $^{27}$Al system in Fig.~\ref{fig1}, where 
the extracted FF and DI components  for Be, B, C and N fragments have been 
displayed (dotted and dash-dotted curves, respectively) alongwith the 
experimental data for $\theta_{lab}  $=  15$^\circ$.
It is clear from Fig.~\ref{fig1} that the experimental energy spectra for all the
fragments are nicely fitted with two Gaussians representing FF and DI 
components. To investigate further the applicability of the scheme over the 
whole angular range of the data, experimental energy spectra of Carbon and Nitrogen
fragments for the same system at  two other angles (20$^\circ$, 40$^\circ$)   have also 
been displayed alongwith the respective estimates of FF and DI components in 
Fig.~\ref{fig2a}. It is clear from the figure that  in these cases too, the 
above scheme is fairly successful in estimating the experimental energy spectra.
For further illustration, experimental energy spectra of Carbon and Nitrogen
fragments at two different angles for the other systems (i.e., $^{16}$O (116 MeV) + 
$^{28}$Si, $^{20}$Ne (145 MeV) + $^{27}$Al, $^{59}$Co)   have been displayed 
along with the respective estimates of FF and DI components in 
Figs.~\ref{fig2b},~\ref{fig2c},~\ref{fig2d}, respectively. It is evident that in all cases the 
above scheme for the decomposition of FF and DI components works  fairly well
in estimating the experimental energy spectra.

\begin{figure}


\centering
{\psfig{figure=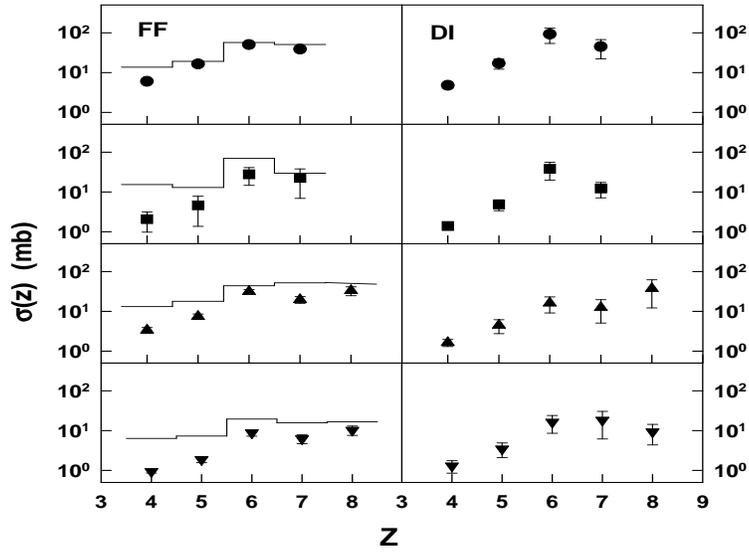,width=10.5cm,height=12.0cm}}


\caption{Variation of total elemental yields, $\sigma (Z)$, of FF (left) and 
DI (right) components, plotted as function of fragment charge, $Z$, 
for different systems. Circle, square, triangle and inverted triangle 
correspond to the experimental estimates of  $\sigma (Z)$ for the reactions 
$^{16}$O (116 MeV) + $^{27}$Al, $^{16}$O (116 MeV) + $^{28}$Si, 
$^{20}$Ne (145 MeV) + $^{27}$Al and $^{20}$Ne (145 MeV) + $^{59}$Co,
respectively. The solid histograms are the corresponding  EHFM  predictions
of the total elemental FF yields.} 
\label{fig2e} 
\end{figure}

\paragraph{Total elemental yields : }
The FF and DI components of the total elemental yields,
extracted using the procedure outlined above, have been displayed in
Fig.~\ref{fig2e} for all the reactions under present study. The FF
components of the fragment emission cross-sections have been compared  with
the  theoretical  estimates  of  the  same  obtained  from  the  extended
Hauser-Feshbach   method   (EHFM)   \cite{Mat97}.   The   values of
the critical angular momentum for fusion, $l_c$
and the grazing angular momentum, $l_g$ for the systems considered here
have been given in Table~\ref{tab1}. 
The values of critical angular momenta have been obtained 
from experimental fusion cross section data, whereever available 
\cite{pap,dje}. Otherwise, they have been obtained from 
dynamical trajectory model calculations with realistic nucleus-nucleus
interaction and dissipative forces generated self-consistently through
stochastic nucleon exchanges \cite{bh2}. The $l_c$ values predicted by the 
dynamical model  have  been cross checked with the respective available 
experimental values and they were found to be in excellent agreement
(e.g., for $^{16}$O  + $^{28}$Si and  $^{20}$Ne + $^{27}$Al systems, predicted
values of $l_c$ were 35$\hbar$ and 37$\hbar$, respectively, which were same
as their respective experimental estimates - see Table~\ref{tab1}).
The calculated fragment emission
cross-sections are shown in Fig.~\ref{fig2e} as solid  histogram  and
compared with the experimental estimates of the same (different symbols 
correspond to different reactions). It
is  seen  from  the  figure  that the theoretical predictions are in fair
agreement with the experimental results. 

\begin{figure}


\centering
{\psfig{figure=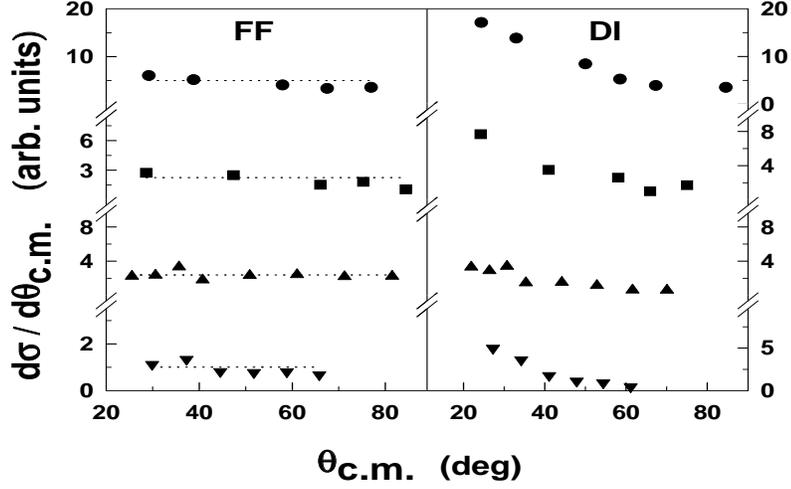,width=10.5cm,height=12.0cm}}


\caption{Variation of cross sections of FF and 
DI components for Carbon fragment, plotted as function
of centre of mass angle, $\theta_{c.m.}$ for different systems. Circle,
square, triangle and inverted triangle correspond to 
$^{16}$O (116 MeV) + $^{27}$Al, $^{16}$O (116 MeV) + $^{28}$Si, 
$^{20}$Ne (145 MeV) + $^{27}$Al, $^{20}$Ne (145 MeV) + $^{59}$Co,
respectively. The dotted curves correspond to fission-like angular
distribution ($ d\sigma / d\Omega \sim a/\sin \theta_{c.m.}$) fit to
the FF component of the data. } 
\label{fig3} 
\end{figure}

\paragraph{Angular distributions : } 
The centre of mass angular distributions of FF and DI components for a typical 
ejectile Carbon emitted in the reactions mentioned above have been displayed in 
Fig.~\ref{fig3} as a function of centre of mass angle  $\theta_{c.m.}$. 
The centre of mass (c.m.) angular distributions of the FF components, 
as expected, are found to be symmetric ($ \propto 1/\sin \theta_{c.m.}$ ), 
whereas those of the DI components are falling off more rapidly indicating
shorter lifetime of the dinuclear composite (Fig.~\ref{fig3}). 

\begin{figure}


\centering
{\psfig{figure=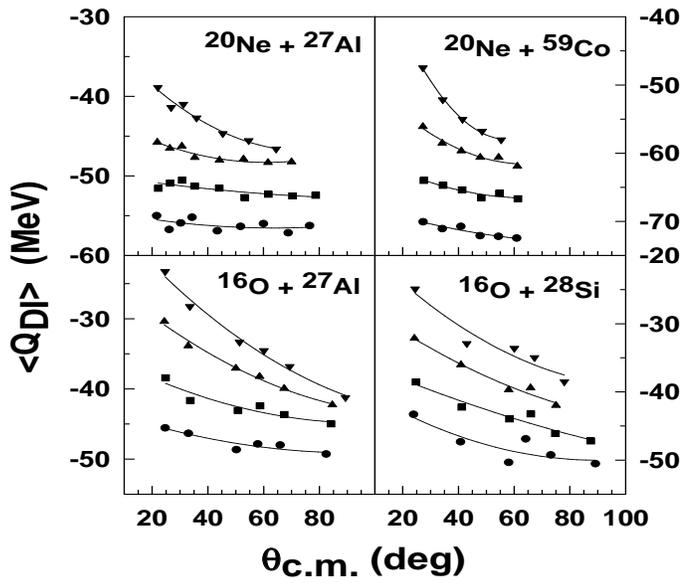,width=10.5cm,height=12.0cm}}


\caption{ Variation of optimum Q-values for deep inelastic reaction, 
$<Q_{DI}>$, plotted as function of centre of mass angle, $\theta_{c.m.}$
for different systems. Circle, square, triangle and inverted triangle correspond
to the fragments Be, B, C and N, respectively. 
Curves are drawn to guide the eye.} 
\label{fig4} 
\end{figure}

\paragraph{Average Q-values : }
The average  Q-values for the DI fragments  ($<Q_{DI}>$) have been displayed
in Fig.~\ref{fig4} as a function of c.m. angle ($\theta_{c.m.}$). The 
Q-values have been estimated from fragment kinetic energies assuming two body
kinematics. The fragment kinetic energies were appropriately corrected for 
particle evaporation from the excited primary fragments assuming thermal 
equilibrium of the dinuclear composite system. The values of 
$<Q_{DI}>$ for Be and B are found to be nearly constant as a function of 
angle, whereas those for C and N are found to decrease at forward angles 
($\theta_{c.m.} \lesssim 40^\circ$) and then the two gradually tend to become 
constant; these imply that, beyond this point, the kinetic energy damping is complete 
and dynamic equilibrium has been
established before the scission of the dinuclear composite takes place. In the
following,  we try to explore these completely damped collisions further to 
extract the magnitude of angular mometum damping.

\section{DISCUSSION}
\label{sec:dis}

\begin{figure}


\centering
{\psfig{figure=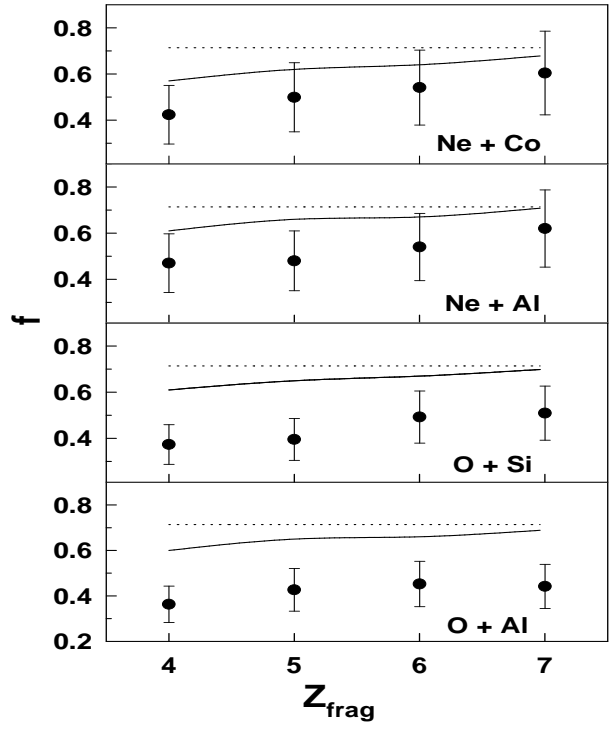,width=10.5cm,height=12.0cm}}


\caption{ Variation of angular momentum dissipation factor $f$ with 
fragment. The filled circles are extracted from the data, solid and dotted  
curves correspond to sticking limit and rolling limit predictions, respectively.} 
\label{fig5} 
\end{figure}

As mentioned earlier, magnitude of angular momentum damping may be estimated
from Eq.~\ref{eq:ef} only if scission configuration can be estimated 
independently. Assuming the friction to be at its limit (sticking limit), the extracted
scission configuration for $^{20}$Ne (120 MeV) + $^{27}$Al was found to be 
$\sim$11 fm \cite{na}, which is much larger than the sum of nuclear radii. 
However, experimental study of DI collision in the reaction
$^{20}$Ne (151 MeV) + $^{27}$Al \cite{se} indicated that scission configuration 
of the fully damped component (at larger angles) may be quite compact, whereas
that for the partially damped component (at smaller angles) may be quite 
elongated having neck length $\sim$3.7 fm. This may be intuitively justified as
follows. Deep inelastic collisions are believed to occur within the angular
momentum window between the critical angular momentum for fusion, $l_c$
and the grazing angular momentum, $l_g$. The partially damped part
of it (at forward angles)  originate in near peripheral collisions  ($l \sim l_g$),
which correspond to small overlap and thus a fairly elongated dinuclear
configuration; on the other hand, fully damped components (at larger angles)
correspond to more compact collisions near $l \sim l_c$. Interestingly, 
fusion-fission yield is also most predominant in the vicinity of $l \sim l_c$. It is,
therefore, likely that the exit channel configurations of both the processes
are similar and  it appears to be fairly
reasonable to assume a compact scission shape for the fully damped 
component of the data. In the present work, we estimated the scission configuration
from the extracted fusion-fission component of the measured fragment 
energy spectra.  The separation distance $d$ between the the two fragments
at the scission point is calculated from the energy centroid of the FF energy spectra 
which obeyed  Viola systematics \cite{vi} corrected for asymmetric mass splitting 
\cite{be96}. The mean values of $d$ thus estimated  are; $7.0  \pm 0.7$ fm for  
$^{16}$O  + $^{27}$Al, $7.2  \pm 0.7$ fm for $^{16}$O  + $^{28}$Si, $7.7  \pm 1.2$ fm for 
$^{20}$Ne  + $^{27}$Al and $10.9  \pm 1.9$ fm for $^{20}$Ne  + $^{59}$Co. 
Assuming these scission configurations corresponding to each mass 
splitting to be 'frozen',  Eq.~\ref{eq:ef} may then be used to extract the mean
angular momentum dissipation factor, $f$, in the case of DI collisions. The
values of $f$ extracted for different systems are displayed in
Fig.~\ref{fig5} (filled circles) alongwith the rolling and sticking limit predictions 
(dotted and solid curves, respectively) for the same. For the purpose of 
evaluation of $f$, the value of 
initial angular momentum $l_i$ was taken to be equal to the critical 
angular momentum for fusion, $l_c$.

It is apparent from Fig.~\ref{fig5} that for all the reactions considered,
there is discrepancy between the experimental and empirical estimates
of angular momentum dissipation. In all cases, the experimental 
estimates of the mean angular momentum dissipation 
are more than their  limiting values predicted by the 
sticking condition (for $^{20}$Ne  + $^{59}$Co reaction, however, 
the experimental estimates of $f$ and the corresponding sticking limit
predictions are within the ranges of experimental uncertainties).
The discrepancy is more for lighter fragments, and
gradually decreases for heavier fragments. This may be intuitively
understood as follows;  it is known from the study of dissipative dynamics 
of fission ( see, for example, Ref.~\cite{dhara} and references therein)
that, strong frictional forces in the exit channel cause
considerable retardation of the scissioning process leading to increase in
scission time scale. As the exit channel configurations of the fully damped 
DI process  are taken to be similar to those for FF process (except that 
the dinuclear system, in case of DI collision,  is formed beyond the conditional 
saddle point directly), the dynamics of DI process may also experience
stronger frictional forces. Microscopically, friction is generated due to
stochastic exchange of nucleons between the reacting partners through
the window formed by the overlap of the density distributions of the two.
Stronger friction, in this scenario, essentially means larger degree of density overlap
and more nucleon  exchange. Consequently,  lighter DI fragments 
(corresponding to more net nucleon transfer) originate
from deeper collisions, for which  interaction times are larger. 
Therefore, angular momentum dissipation too,  originating due to
stochastic nucleon exchange, may  be more 
which, at least qualitatively, explains the observed trend. 
Moreover, it is also seen that the difference 
between the experimental estimates and the corresponding sticking limit 
predictions is more for lighter systems ($^{16}$O + $^{27}$Al, $^{28}$Si, 
$^{20}$Ne + $^{27}$Al), and less for the heavier system ($^{20}$Ne + $^{59}$Co).
Qualitatively, this may be due to entrance channel effect \cite{sza96}; as the formation 
time (of shape equilibrated fused composite) is smaller for the lighter system at lower
spin, the two time scales (of fusion and DI processes) are closer. This may give
rise to larger angular momentum dissipation for lighter systems as observed in
the present work.

\section{SUMMARY AND CONCLUSIONS}
\label{sec:sum}
In summary, we have studied fragment emission from $^{16}$O (116 MeV) + $^{27}$Al, 
$^{28}$Si, $^{20}$Ne (145 MeV) + $^{27}$Al, $^{59}$Co reactions and extracted
the contributions of fusion-fission and deep inelastic components.  Assuming
a compact exit channel configuration for the fully damped part of the DI reactions,
the exit channel configuration has been estimated from the extracted FF part of
the spectra. The angular momentum dissipation for the fully damped DI reactions has 
then been extracted using these scission shapes. The angular momentum dissipations
have been found to be more than the corresponding sticking limit predictions in all the 
cases except for the case of $^{20}$Ne + $^{59}$Co, where, the mean values of the
experimental estimate of angular momentum dissipation are systematically less than the
corresponding sticking limit values, though they are within the range of experimental uncertainty.
 This may be due to stronger friction in the exit channel which may cause
longer overlap of the dinuclear system and consequently more nucleon exchange and 
dissipation of angular momentum due to stochastic nature of nucleon exchange. 
The effect is more for lighter systems, as in this case there  is more overlap in the time scales
of FF and DI processes. However, further systematic studies for each system at different 
bombarding energies are needed for a better
understanding on the dissipation mechanism in light nuclear systems. 
The inclusive yields for some fragments may have additional contribution from other reaction 
mechanisms like projectile breakup process (e.g., $\alpha$-breakup in the case of Ne 
projectile), which should be properly taken care of. The inclusive data
presented in this paper may also be useful for future exclusive experiments for which 
light charged particles are detected in coincidence with either fully damped deep-inelastic
or fusion-fission fragments.

\acknowledgements

The authors like to thank cyclotron operating staff for smooth running of the machine, 
R. Saha and H. P. Sil for the fabrication of thin Si detectors for the experiment.
One of the authors (C.B) likes to thank C. Beck of IReS, Strasbourg, for 
valuable discussions. One of the authors (A. D)  acknowledges with thanks  the financial 
support received from C. S. I. R., Government of India.

\newpage
\begin{table}
\caption{Values of critical and grazing angular momenta ($l_c$, $l_g$).}   
\begin{tabular}{ccc}
Reaction&$l_c$&$l_g$\\
\tableline
$^{16}$O + $^{27}$Al&34\tablenotemark[1]&43\\

$^{16}$O + $^{28}$Si&35\tablenotemark[2]&44\\

$^{20}$Ne + $^{27}$Al&37\tablenotemark[3]&50\\

$^{20}$Ne + $^{59}$Co&55\tablenotemark[1]&63\\
\end{tabular}
\tablenotetext[1]{From theoretical calculation, Ref.\ \cite{bh2}.}
\tablenotetext[2]{From experimental fusion data, Ref.\ \cite{pap}.}
\tablenotetext[3]{From experimental fusion data, Ref.\ \cite{dje}.}
\label{tab1}
\end{table}

\end{document}